\documentclass[fleqn,12pt]{article}
\usepackage[numbers,sort&compress]{natbib}
\usepackage[footnotesep=0.4in]{geometry}
\usepackage{graphicx}
\usepackage{caption2}
\usepackage{amsmath}
\usepackage{bbding}
\usepackage{graphics}
\usepackage{pifont}
\usepackage{amsfonts}
\usepackage{caption2}
\usepackage{float}
\makeatletter
\newcommand\figcaption{\def\@captype{figure}\caption}
\newcommand\tabcaption{\def\@captype{table}\caption}
\DeclareMathOperator{\sech}{Sech}
 \topmargin
-0.8in \textheight 9.4in \textwidth6.8in \hoffset -1.45cm

\begin{document}

\date{}
\title{Solitons and breathers for nonisospectral mKdV equation with Darboux transformation}
\author{Ling-Jun Liu,   Xin Yu\thanks{Corresponding
author, with e-mail address as yuxin@buaa.edu.cn}
\\{\em Ministry-of-Education Key Laboratory of Fluid Mechanics and
National}\\
{\em  Laboratory for Computational Fluid Dynamics,  Beijing University of }\\
{\em  Aeronautics and Astronautics, Beijing 100191, China} }

\maketitle

\vspace{8mm}

\begin{abstract}

\vspace{3mm} Under investigation in this paper is the nonisospectral
and variable coefficients modified Kortweg-de Vries (vc-mKdV)
equation, which manifests in diverse areas of physics such as fluid
dynamics, ion acoustic solitons and plasma mechanics. With the
degrees of restriction reduced, a simplified constraint is
introduced, under which the vc-mKdV equation is an integrable system
and the spectral flow is time-varying. The Darboux transformation
for such equation is constructed, which gives rise to the generation
of variable kinds of solutions including the double-breather
coherent structure, periodical soliton-breather and localized
solitons and breathers. In addition, the effect of variable
coefficients and initial phases is discussed in terms of the soliton
amplitude, polarity, velocity and width, which might provide
feasible soliton management with certain conditions taken into
account.

\end{abstract}


\vspace{3mm}


\vspace{20mm}

\newpage

\noindent {\Large{\bf I. Introduction}}

 \vspace{3mm}

The modified Kortweg-de Vries (mKdV) type equations arise in diverse
areas of physics, including fluid dynamics, ion acoustic solitons
and plasma mechanics~\cite{fzt,kh,hl,dkg,wyl}. For instance, the
dynamics of the interfacial waves in a two-layer fluid of slowly
varying depth is studied by the formulated standard mKdV
equation~\cite{kh}, the ion-acoustic solitary waves in certain
unmagnetized plasma is described by the cylindrical and spherical
mKdV equation~\cite{dkg} and the propagation of circularly polarized
few-cycle pulses with wave polarization is investigated by a
non-integrable complex mKdV equation~\cite{hl}. With the
inhomogeneous environmental density and boundary conditions taken
into account~\cite{rg}, the constant coefficients mKdV equation can
be extended as followings:
\begin{equation}\label{equation}
\hspace{10mm}u_t\,+a(t)\,u^2\,u_{x}+b(t)\,u_{xxx}+[c_0(t)+c_1(t)x]\,u_{x}+d(t)\,u=0\,,
\end{equation}
where a(t), b(t), $c_0(t), c_1(t)$ and d(t) are functions of
variable t.

With different selections of coefficients, Eq.~(\ref{equation}) has
been investigated for physical interest~\cite{zy,ov,gxl,pde}.
Thereinto, periodic wave solutions are constructed in bilinear forms
based on a multidimensional Riemann theta function~\cite{zy}, group
classification is carried out and all the classes under
consideration are normalized~\cite{ov}, the pulse waves in thin
walled prestressed elastic blood vessels is described~\cite{gxl} and
a modified perturbation technique is applied to the problem of the
structural instability of algebraic solitons~\cite{pde}.

For constructing soliton solutions from the soliton equations like
Eq.~(\ref{equation}), there exist many methods such as Hirota
bilinear method~\cite{szy,wcj,zy2}, associated scattering
problem~\cite{ep} and Darboux transformation
(DT)~\cite{zql1,zql2,wl,gxl2,jlj}, in which the DT has proven itself
a purely algebraic iterative tool~\cite{zql1,zql2}. The main
difficulty in finding a proper spectral problem of constructing DT
has been investigated and many spectral problems have been searched
for, including different forms of DT of Eq.~(\ref{equation}) under
certain conditions~\cite{wl,zql2,gxl2,jlj}.

Except for soliton solutions, Eq.~(\ref{equation}) also has
solutions in the form of oscillating packets (breathers)~\cite{pde},
which along with the solitons, determine the asymptotics of the wave
field~\cite{AVS}. It is believed that some temporal and spatial
variability that has been observed in oceanic internal soliton
fields may be due to the breather interaction~\cite{ckw} and several
investigations have been carried out~\cite{ckw,sc,lkg,AVS}. The
circumstances supporting the formation of breathers are determined
with various piecewise-constant initial conditions~\cite{sc}, some
detailed examinations of the breather-soliton interaction process
are analyzed by the Hirota bilinear method~\cite{ckw} and the
breather in numerical simulations using the full nonlinear Euler
equations for stratified fluid is presented under several forms of
the initial disturbance~\cite{lkg}.

Meanwhile, Eq.~(\ref{equation}) becomes integrable when the variable
coefficients satisfy certain constraint conditions, in which
$a(t)=6\,b(t)$ and $d(t)=c_1(t)$, leading to analytical discussion
related to different branches of
physics~\cite{wlchan1,wlchan2,szy,dcq,fq,zs,yzy}. Different from
above investigation, hereby we introduce a constraint reducing the
degrees of restriction
\begin{equation}\label{relation}
\hspace{10mm}a(t)=6\,b(t)\,e^{\int{\!2\,[d(t)-c_1(t)]dt}}\,,
\end{equation}
under which Eq.~(\ref{equation}) is an integrable system and there
can be an integrable constant which is normalized to be 1 in this
case.

However, to our best knowledge, the Lax pair and DT for
Eq.~(\ref{equation}) under constraint~(\ref{relation}) have not been
obtained. Therefore, in this paper, we will construct the DT with
spectral problem for Eq.~(\ref{equation}) and generate multi-soliton
and breather solutions from the seed ones.

The outline of the present paper is as follows. In Section II and
III, the Lax pair and DT for Eq.~(\ref{equation}) under
constraint~(\ref{relation}) are constructed respectively. In Section
IV, one-soliton solution is generated by a DT and the characteristic
line with soliton amplitude is given. In Section V, two-soliton and
three-soliton solutions are investigated analytically by appropriate
selection of variable coefficients. In Section VI, different kinds
of breathers as well as the interaction between breather and soliton
are thrived, including the double-breather coherent structure,
periodical soliton-breather and localized breathers. In Section VII,
the conclusions are given.

\vspace{7mm} \noindent {\Large{\bf II. Lax pair}}

\vspace{3mm} Based on AKNS procedure~\cite{gxl2}, the Lax pair of
Eq.~(\ref{equation}) under the general constraint~(\ref{relation})
is constructed with the undetermined coefficients corresponding to
time and space

\begin{equation}\label{akns1}
\Phi_x(x,t)=U\Phi(x,t)= \left(
  \begin{array}{ccc}
    \lambda(t) & l(t)\,u\\
    -l(t)\,u & -\lambda(t)\\
  \end{array}
\right)\Phi(x,t),
\end{equation}

\begin{equation}\label{akns2}
\Phi_t(x,t)=V\Phi(x,t)=\left(
  \begin{array}{ccc}
    A(x,t) & B(x,t)\\
    S(x,t) & -A(x,t)\\
  \end{array}
\right)\Phi(x,t),
\end{equation}
where $\Phi(x,t)=(\Phi_1(x,t),\Phi_2(x,t))^\textit{T}\,$ with
\textit{T }representing the transpose of the vector. The $\Phi(x,t)$
should comply with the compatibility condition, which leads to a
zero curvature equations~\cite{zql1}

\begin{equation}\label{zerocurvature}
\hspace{10mm}U_x-V_t+[U,V]=0.
\end{equation}

Substituting Eqs.~(\ref{akns1}) and~(\ref{akns2}) into the zero
curvature equations and comparing the coefficients of the power of
$\lambda(t)$, we have

\begin{equation}\label{l}
\hspace{10mm}l(t)=e^{\int{\![d(t)-c_1(t)]dt}},
\end{equation}

\begin{equation}\label{A}
\hspace{10mm}A(x,t)=-\big(2\,e^{2\,\int{\![d(t)-c_1(t)]dt}}\,b(t)\,u^2+c_0(t)+x\,c_1(t)\big)\,\lambda(t)-4\,b(t)\,\lambda^3(t),
\end{equation}

\begin{equation}\label{B}
\begin{aligned}
\hspace{10mm}B(x,t)=
-e^{\int{\![d(t)-c_1(t)]dt}}\,\big(2\,e^{2\,\int{\![d(t)-c_1(t)]dt}}\,b(t)\,u^3+c_0(t)\,u
\\
+x\,c_1(t)\,u+4\,b(t)\,\lambda^2(t)\,u+2\,b(t)\,\lambda(t)\,u_x+b(t)\,u_{xx}\big)\,,
\end{aligned}
\end{equation}

\begin{equation}\label{C}
\begin{aligned}
\hspace{10mm}C(x,t)=
e^{\int{\![d(t)-c_1(t)]dt}}\,\big(2\,e^{2\,\int{\![d(t)-c_1(t)]dt}}\,b(t)\,u^3+c_0(t)\,u
\\
+x\,c_1(t)\,u+4\,b(t)\,\lambda^2(t)\,u-2\,b(t)\,\lambda(t)\,u_x+b(t)\,u_{xx}\big)\,.
\end{aligned}
\end{equation}

Meanwhile, spectral parameter $\lambda(t)$ should satisfy the
following relation
\begin{equation}\label{lambdarelation}
\hspace{10mm}\lambda'(t)+c_1(t)\,\lambda(t)=0.
\end{equation}

\vspace{7mm} \noindent {\Large{\bf III. Construction of DT}}

Now we introduce a gauge transformation for the spectral problem
\begin{equation}\label{gt}
\hspace{10mm} \overline{\Phi}=M\,\Phi\,,
\end{equation}
where $M$ is defined by
\begin{equation}\label{reladt1}
\hspace{10mm} M_x+M\,U=\overline{U}\,M
\end{equation}
\begin{equation}\label{reladt2}
\hspace{10mm} M_t+M\,V=\overline{V}\,M
\end{equation}

The gauge transformation is called DT~\cite{wl}, if the spectral
problem~(\ref{akns1}) and~(\ref{akns2}) can be transformed into
\begin{equation}\label{akns21}
\overline{\Phi}_x=\overline{U}\,\overline{\Phi},
\end{equation}
\begin{equation}\label{akns22}
\overline{\Phi}_t=\overline{V}\,\overline{\Phi},
\end{equation}
where $\overline{V}$ and $\overline{U}$ has the same form as U and V
and the old potential $u$ is mapped into a new potential $u_1$.

Suppose
\begin{equation}\label{DT}
M=\,m(t)\,\lambda(t)\left(
  \begin{array}{ccc}
    1 & 0\\
    0 & 1\\
  \end{array}
\right)+
 n(t)\,\lambda_1(t)\left(\begin{array} {ccc}
    a_0(x,t) & b_0(x,t)\\
    c_0(x,t) & d_0(x,t)\\
  \end{array}
\right),
\end{equation}
where $\lambda_1(t)=e^{\int{\![-c_1(t)]dt}}\,\overline{\lambda_1}$.

Substituting Expression~(\ref{DT}) into Eqs.~(\ref{akns21})
and~(\ref{akns22}), by a direct calculation, we have
\begin{equation}\label{m}
\hspace{10mm} m(t)=n(t)=e^{\int{\![c_1(t)]dt}}\,,
\end{equation}
\begin{equation}\label{ad0}
\hspace{10mm}
a_0(x,t)=-d_0(x,t)=\frac{f_{2}^2-f_{1}^2}{f_{1}^2+f_{2}^2}\,,
\end{equation}
\begin{equation}\label{bc0}
\hspace{10mm}
b_0(x,t)=c_0(x,t)=-\frac{2\,f_{1}\,f_{2}}{f_{1}^2+f_{2}^2}\,,
\end{equation}
where
$\Big(f_1\big(x,t,\lambda_1(t)\big),f_2\big(x,t,\lambda_1(t)\big)\Big)^\textit{T}=\Big(\Phi_1(x,t),\Phi_2(x,t)\Big)^\textit{T}\,$.

The transformation between old potentials $u$ and new ones $u_1$ is
given as below
\begin{equation}\label{u1}
\hspace{10mm}
u_1(x,t)=u(x,t)+4\,e^{\int{\![c_1(t)-d(t)]dt}}\,\lambda_1(t)\,\frac{f_{1}\,f_{2}}{f_{1}^2+f_{2}^2}\,.
\end{equation}

So far, we have obtained the DT~(\ref{DT})-(\ref{u1}), which
transforms the matrix spectral problem into another spectral problem
of the same type and can generate new solutions from seed ones by
purely algebraic iteration.

It is worth noting with certain modification, the DT can be extended
as following
\begin{equation}\label{DTguang}
T=\,e^{\int{\![c_1(t)]dt}}\,\lambda(t)\left(
  \begin{array}{ccc}
    1 & 0\\
    0 & 1\\
  \end{array}
\right)-e^{\int{\![c_1(t)]dt}}\,\frac{\lambda_1(t)}{1+\sigma^2}\left(\begin{array}
{ccc}
    1-\sigma^2 & 2\,\sigma\\
    2\,\sigma & -1+\sigma^2\\
  \end{array}
\right)\,,
\end{equation}
\begin{equation}\label{u1guang}
\hspace{10mm}
u_1(x,t)=u(x,t)+4\,e^{\int{\![c_1(t)-d(t)]dt}}\,\lambda_1(t)\frac{\sigma}{1+\sigma^2}\,,
\end{equation}
\begin{equation}\label{sigma}
\hspace{10mm}
\sigma=\frac{f_{22}+\mu_1\,f_{21}}{f_{12}+\mu_1\,f_{11}}\,,
\end{equation}
where
$\Big(f_{11}\big(x,t,\lambda_1(t)\big),f_{21}\big(x,t,\lambda_1(t)\big)\Big)^\textit{T}$
and
$\Big(f_{12}\big(x,t,\lambda_1(t)\big),f_{22}\big(x,t,\lambda_1(t)\big)\Big)^\textit{T}$
are linearly independent solutions of spectral problems.

\vspace{7mm} \noindent {\Large{\bf IV. One soliton solution}}

\vspace{3mm}

Employing $u=0$ as the original solution for Eq.~(\ref{equation}),
we can obtain the solutions of spectral problems~(\ref{akns1})
and~(\ref{akns2}) as below
\begin{equation}\label{f1}
\hspace{10mm}\Phi_1(x,t)=e^{\Lambda(x,t)}\,,
\end{equation}
\begin{equation}\label{f2}
\hspace{10mm}\Phi_2(x,t)=e^{-\Lambda(x,t)}\,,
\end{equation}
\begin{equation}\label{f3}
\hspace{10mm}\Lambda(x,t)=\lambda(t)\,x-\int{\![c_0(t)\,\lambda(t)+4\,b(t)\,\lambda^3(t)]dt}-\xi\,,
\end{equation}
where $\xi$ is an integration constant representing soliton initial
phase, in every iteration which can change to a certain value.

For a given spectral parameters $\lambda_1(t)$ and initial phase
$\xi_1$ , the explicit solutions
$\Big(f_1\big(x,t,\lambda_1(t)\big),\\f_2\big(x,t,\lambda_1(t)\big)\Big)^\textit{T}$
can be obtained in terms of Expressions~(\ref{f1})-(\ref{f3}), which
leads to a new one soliton solution with the aid of DT~(\ref{u1})
\begin{equation}\label{u_1}
\hspace{10mm}u_1(x,t)=2\,e^{\int{[-d(t)+c_1(t)]dt}}\,\lambda_1(t)\sech{\big[2\,\Omega(x,t)\big]}\,,
\end{equation}
where
\begin{equation}\label{T}
\hspace{5mm}\Omega(x,t)=k(t)\,x-w(t)+\xi_1\,,
\end{equation}
\begin{equation}\label{w}
\hspace{5mm}k(t)=\lambda_1(t)\,,
\end{equation}
\begin{equation}\label{k}
\hspace{5mm}w(t)=\lambda_1(t)\,e^{\int{c_1(t)dt}}\,\int{\big[\,e^{-\int{c_1(t)dt}}(4\,\lambda_1^2(t)\,b(t)+c_0(t))\,\big]dt}\,,
\end{equation}

Meanwhile, the soliton amplitude and characteristic line can be
respectively derived as
\begin{equation}\label{A}
\hspace{5mm}A=2\,e^{\int{[-d(t)+c_1(t)]dt}}\,\lambda_1(t)=2\,\overline{\lambda_1}\,e^{\int{-d(t)dt}}\,,
\end{equation}
\begin{equation}\label{line}
\hspace{5mm}k(t)\,x-w(t)+\xi_1=0\\,
\end{equation}
which indicates one soliton amplitude is only affected by $d(t)$
when $\overline{\lambda_1}$ is given. It is worth noting that the
polarity of soliton also refers to the sign of A. Meanwhile, the
soliton velocity can be obtained by the derivation of
Expression~(\ref{line}).

\vspace{7mm} \noindent {\Large{\bf V. Multi-soliton solutions}}

\vspace{3mm} For general application in various fields related to
soliton dynamics, we need to consider two soliton interaction, which
can be generated by a second DT based on the one-soliton solution.
For soliton characters describing their physical features, soliton
polarity (or phase) should be taken into account. It is worth
mentioning that two soliton polarities are allowed in the mKdV
framework due to the isotropic nonliearity~\cite{AVS} and the
bipolar soliton interaction plays a role in obtaining soliton
management. Therefore, with a positive coefficient $a(t)$ of the
cubic nonlinear term in Eq.~(\ref{equation}), the effect of initial
phase on the soliton polarities will be discussed in the following
section.

\begin{minipage}{\textwidth}
\renewcommand{\captionfont}{\scriptsize}
\renewcommand{\captionlabelfont}{\scriptsize}
\renewcommand{\captionlabeldelim}{.\,}
\renewcommand{\figurename}{Fig.\,}
\hspace{0.5cm}\includegraphics[scale=0.65]{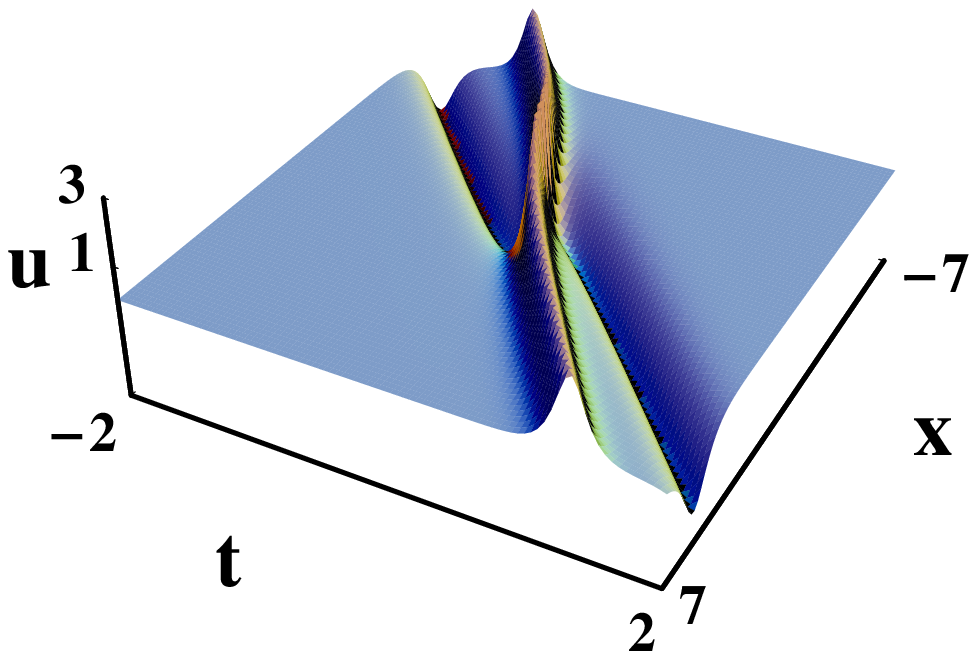}
\hspace{1.5cm}\includegraphics[scale=0.65]{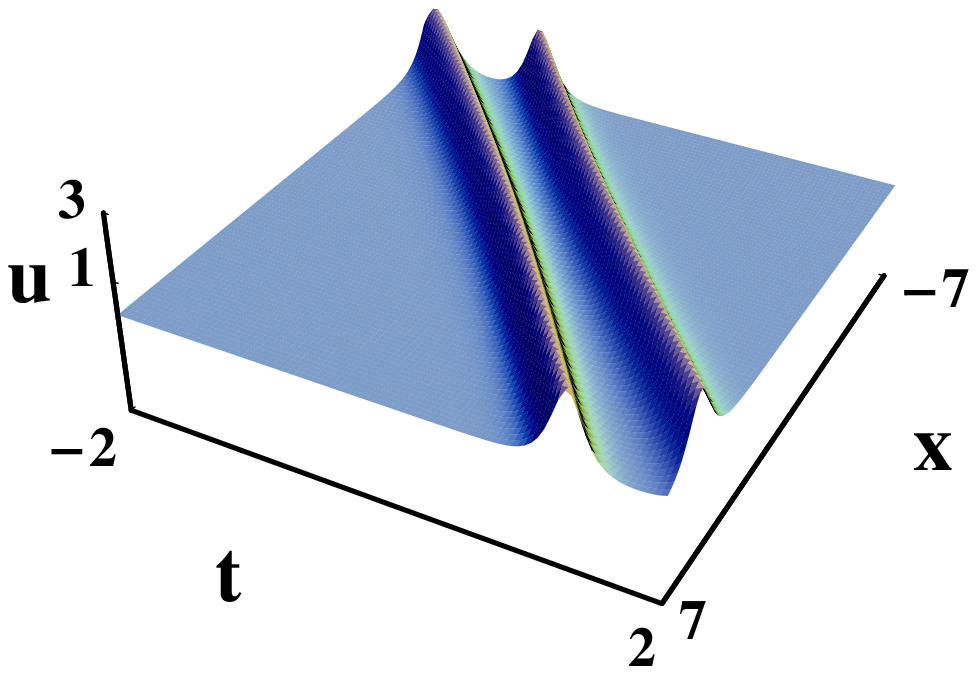}
\vspace{-0.5cm}{\center\hspace{3.3cm}\footnotesize ($a$)
\hspace{7.7cm}($b$)}\figcaption{Solitonic propagation and
interaction with coefficients $\overline{\lambda_1}=1$,
$\overline{\lambda_2}=1.1$, $b(t)=1$, $c_1(t)=c_0(t)=0$, $d(t)=0$
and initial phase: (a) $\xi_1=1,\xi_2=1$, (b)
$\xi_1=1+i\,\frac{\pi}{2},\xi_2=1$} \label{i}
\end{minipage}
\\[\intextsep]

Compared with Fig.~\ref{i}(a), an initial phase shift value of
$i\,\frac{\pi}{2}$ takes place along with the soliton inverse
polarity in Fig.~\ref{i}(b). The depression, which has a negative
amplitude, inverses its polarity while the elevation having the
positive amplitude remains unchanged. As a result, the group changes
from the bipolar solitons into unipolar solitons. The soliton cross
disappears after the inverse polarity of the depression.

\begin{minipage}{\textwidth}
\renewcommand{\captionfont}{\scriptsize}
\renewcommand{\captionlabelfont}{\scriptsize}
\renewcommand{\captionlabeldelim}{.\,}
\renewcommand{\figurename}{Fig.\,}
\hspace{0.5cm}\includegraphics[scale=0.65]{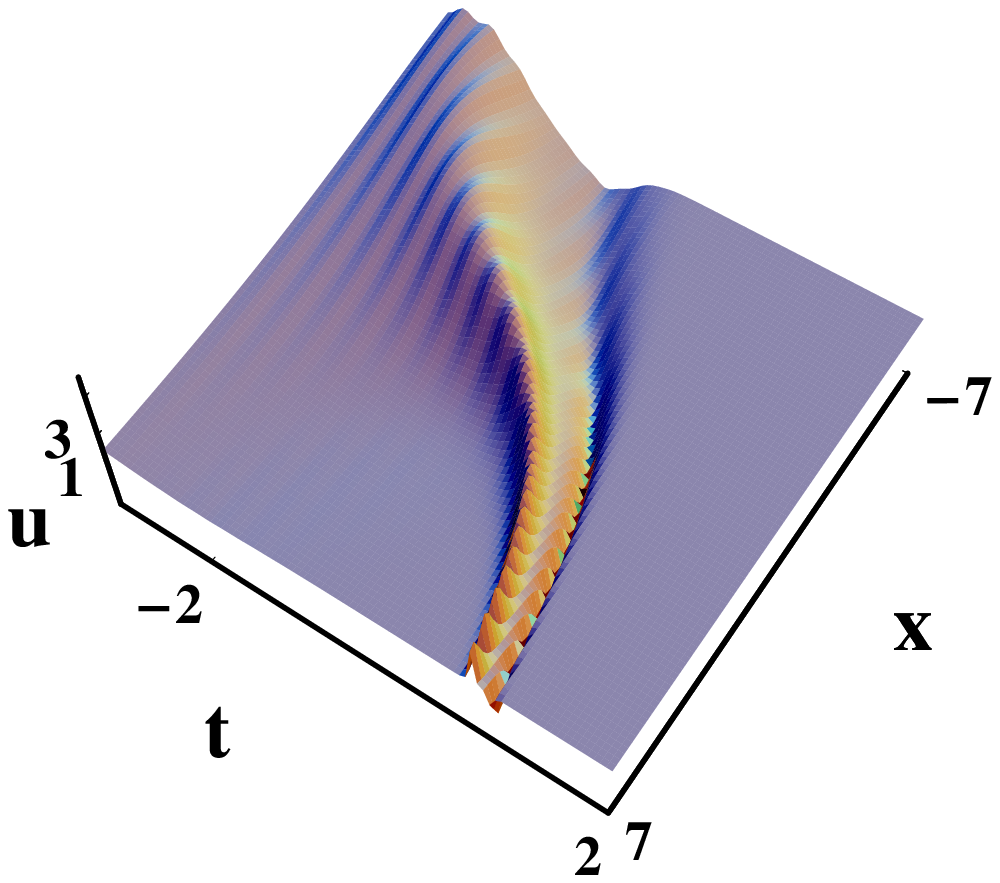}
\hspace{1.5cm}\includegraphics[scale=0.65]{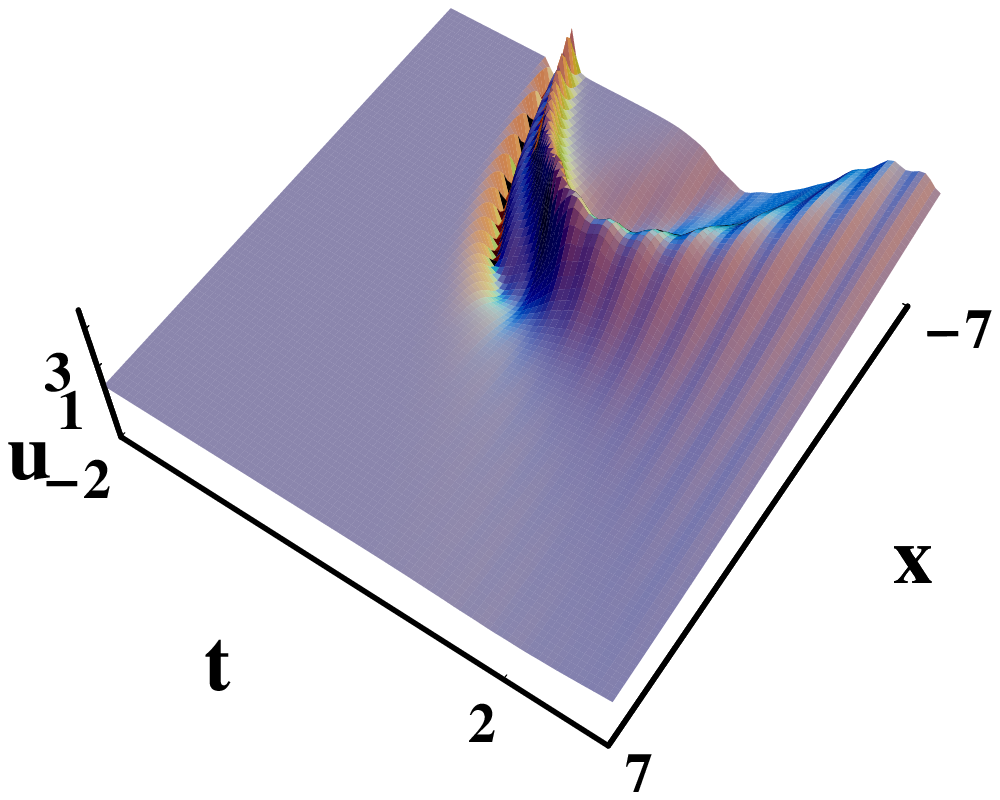}
\vspace{-0.5cm}{\center\hspace{3.3cm}\footnotesize ($a$)
\hspace{7.9cm}($b$)}\figcaption{Solitonic propagation and
interaction with variable coefficients and spectral parameters
$\overline{\lambda_1}=1$, $\overline{\lambda_2}=1.1$,
$\xi_1=1,\xi_2=1$, $b(t)=1$, $c_0(t)=0$, $d(t)=Sin(20t)$ and
line-damping coefficient: (a) $c_1(t)=-1$, (b) $c_1(t)=1$} \label{c}
\end{minipage}
\\[\intextsep]

Under the analytical discussion before, soliton amplitude can be
interpreted by the effect of line-damping coefficient $d(t)$ when
$\overline{\lambda}$ is given. As demonstrated in Fig.~\ref{c}, a
periodic value independent to time is applied for $d(t)$ and
correspondingly the soliton dynamics presents amplitude periodicity.
Coefficient $c_1(t)$ is capable of describing nonuniformity of media
and should be evaluated in the nonisospectral problems. For positive
and negative implication of $c_1(t)$, the solitons propagation
dynamics differs in terms of its width. With $c_1(t)$ being 1 and -1
in this case, the soliton compress and swell respectively along the
propagation.

\begin{minipage}{\textwidth}
\renewcommand{\captionfont}{\scriptsize}
\renewcommand{\captionlabelfont}{\scriptsize}
\renewcommand{\captionlabeldelim}{.\,}
\renewcommand{\figurename}{Fig.\,}
\hspace{0.5cm}\includegraphics[scale=0.65]{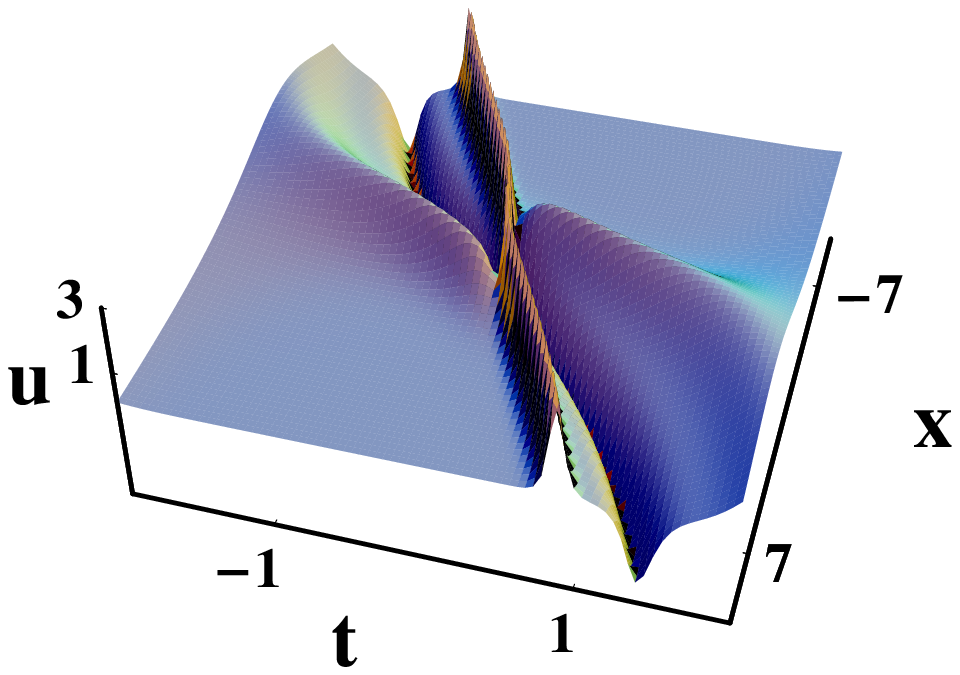}
\hspace{1.5cm}\includegraphics[scale=0.65]{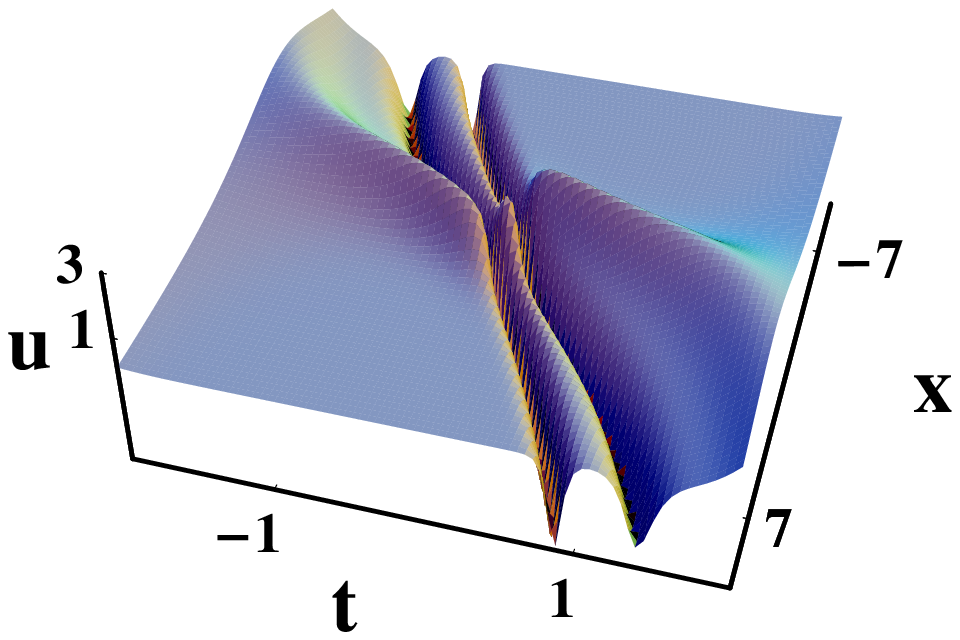}
\vspace{-0.5cm}{\center\hspace{3.3cm}\footnotesize ($a$)
\hspace{7.7cm}($b$)}
\\
\hspace{0.5cm}\includegraphics[scale=0.65]{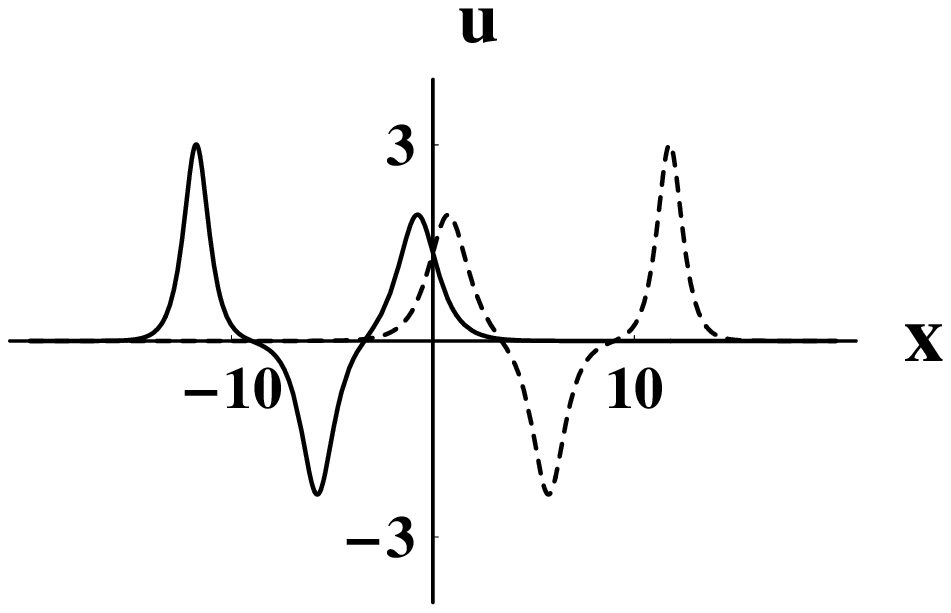}
\hspace{1.5cm}\includegraphics[scale=0.65]{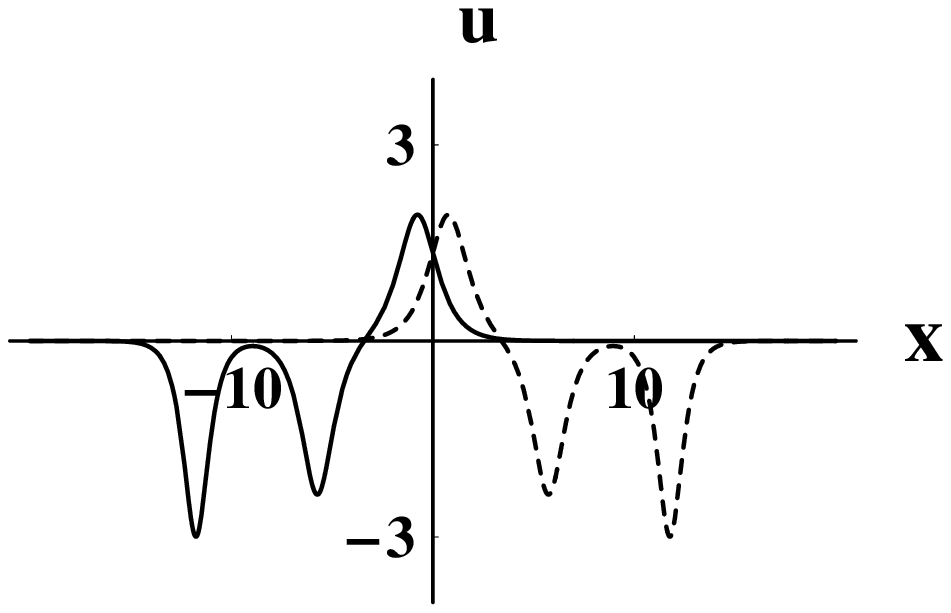}
\vspace{-0.5cm}{\center\hspace{3.3cm}\footnotesize ($c$)
\hspace{7.7cm}($d$)} \figcaption{Interaction among elevation and
depression with variable coefficients and spectral parameters
$b(t)=1$, $c_1(t)=t$, $c_0(t)=0$, $d(t)=0$, $\xi_1=\xi_2=\xi_3=0$,
$\overline{\lambda_1}=1$, and $\overline{\lambda_2}=1.1$,
 (a) $\overline{\lambda_3}=1.5$, (b) $\overline{\lambda_3}=-1.5$. (c)
and (d) profile (a) and (b) respectively when t=-1(solid line) and
t=1(dash line)} \label{sanguzi}
\end{minipage}
\\[\intextsep]

Fig.~\ref{sanguzi} depicts three soliton solution, including the
interaction among depression and elevation. Comparing
Fig.~\ref{sanguzi}(a) with (b), in addition to change the value of
initial phase, we could also manage to replace the elevation by
depression via a sign reversal of spectral parameter
$\overline{\lambda_2}$ from positive to negative. The three soliton
propagation is exhibited by profiles in Figs.~\ref{sanguzi}(c) and
(d) at certain time, with high-amplitude elevation (depression)
firstly leaving behind the low-amplitude elevation (depression) and
then surpassing when time is approaching to 1. Such phenomena are
also indicated in ref~\cite{szy} with Eq.~(\ref{equation}) under an
extra \big($c_1(t)=d(t)$\big).

 \vspace{7mm} \noindent {\Large{\bf VI.
Breathers and breather-soliton solutions}}

\vspace{3mm}Breather solutions can be constructed by DT under a pair
of complex conjugate spectral parameters
($\overline{\lambda_1}=\alpha+i\beta,\overline{\lambda_2}=\alpha-i\beta$).
Similar to the definition that breather can be regarded as a central
valley with two small hills of elevation adjacently, which should
reverse its polarities every half a cycle later~\cite{ckw}, the
breather generated by DT can be also regarded as valley-hill feature
in symmetric form at certain time.

\begin{minipage}{\textwidth}
\renewcommand{\captionfont}{\scriptsize}
\renewcommand{\captionlabelfont}{\scriptsize}
\renewcommand{\captionlabeldelim}{.\,}
\renewcommand{\figurename}{Fig.\,}
\hspace{0.5cm}\includegraphics[scale=0.85]{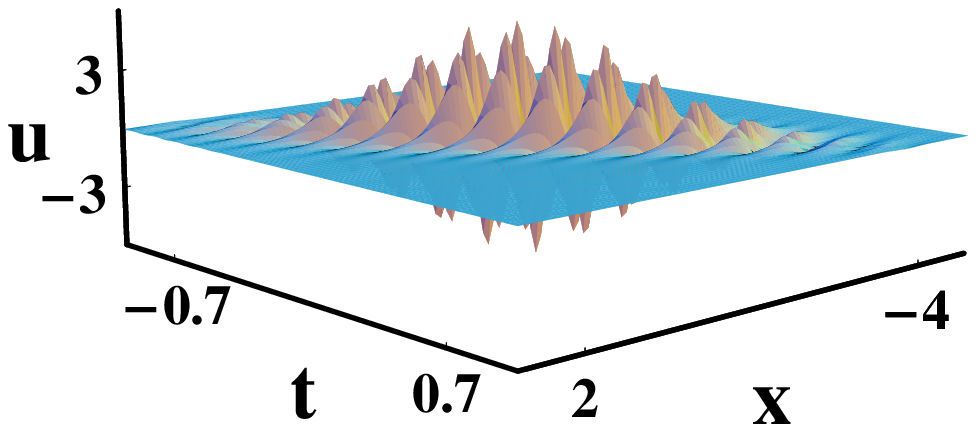}
\hspace{1.5cm}\includegraphics[scale=0.55]{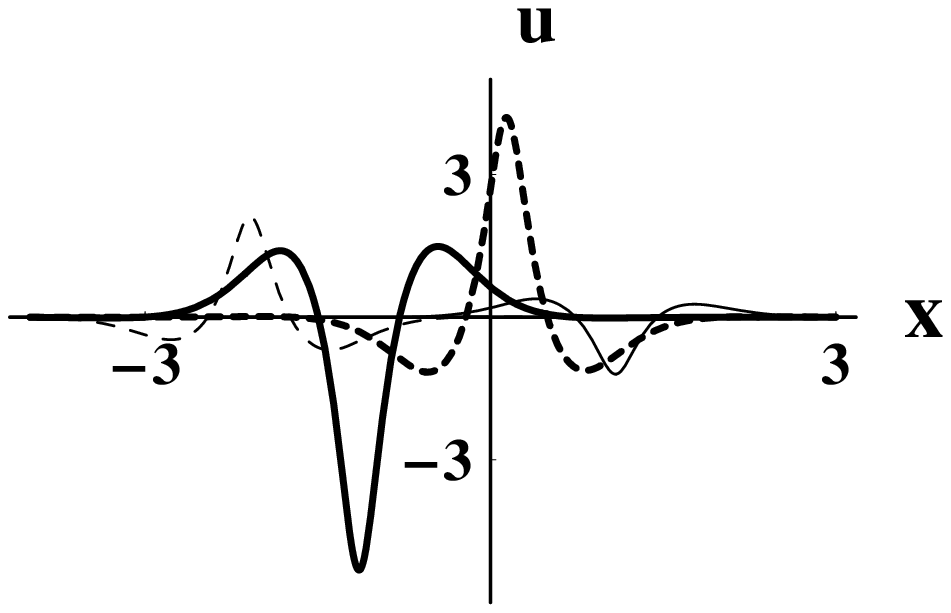}
\vspace{-0.5cm}{\center\hspace{5.3cm}\footnotesize ($a$)
\hspace{7.5cm}($b$)}\figcaption{(a) Breather propagation
$\overline{\lambda_1}=1.5-i$, $\overline{\lambda_2}=1.5+i$,
$\xi_1=\xi_2=1$, $b(t)=1$, $c_0(t)=c_1(t)=0$ and $d(t)=10\,t$; (b)
Profiles (a) when t=-0.566(solid line), t=-0.267(bold dashed line),
t=0.155(bold solid line) and t=0.454(dashed line) }\label{breather}
\end{minipage}
\\[\intextsep]

As illustrated in Fig.~\ref{breather}, the propagation of breathers
demonstrates periodically pulsating and isolated wave forms by
profiles at certain time. Fig.~\ref{breather}(b) illustrates a
downward displacement at center and the horizontal symmetry of
breather structure. With line-damping coefficient $d(t)$ taken into
consideration, breather propagation presents a character of
time-space locality.

\begin{minipage}{\textwidth}
\renewcommand{\captionfont}{\scriptsize}
\renewcommand{\captionlabelfont}{\scriptsize}
\renewcommand{\captionlabeldelim}{.\,}
\renewcommand{\figurename}{Fig.\,}
\hspace{0.5cm}\includegraphics[scale=0.65]{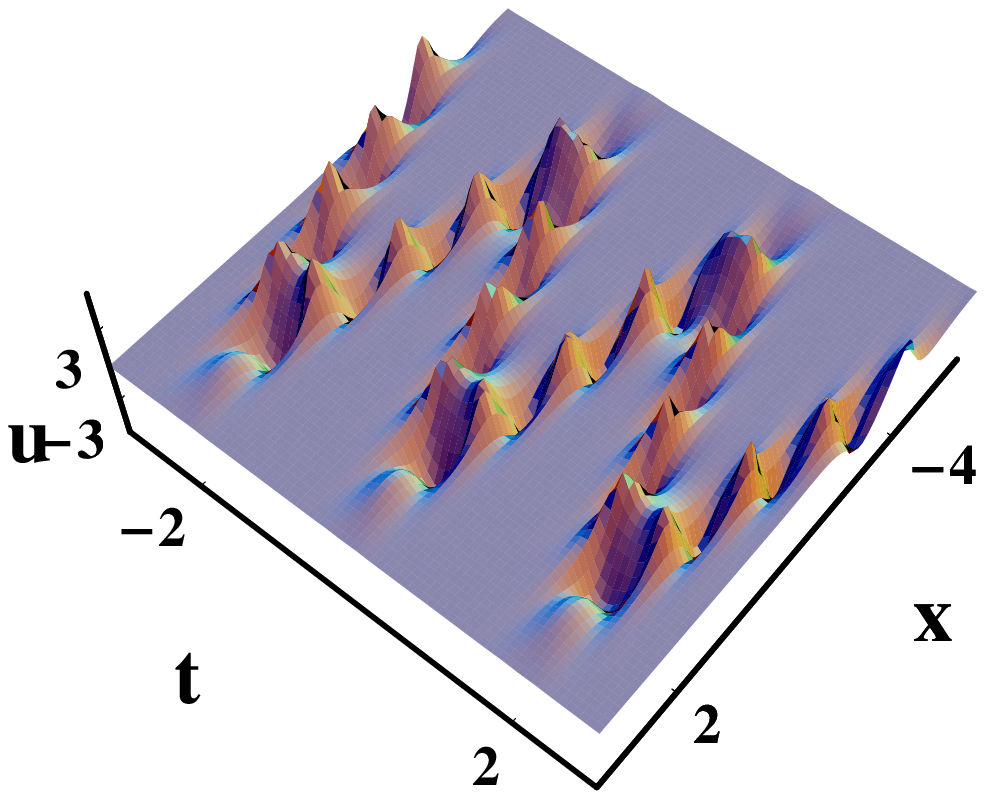}
\hspace{1.5cm}\includegraphics[scale=0.55]{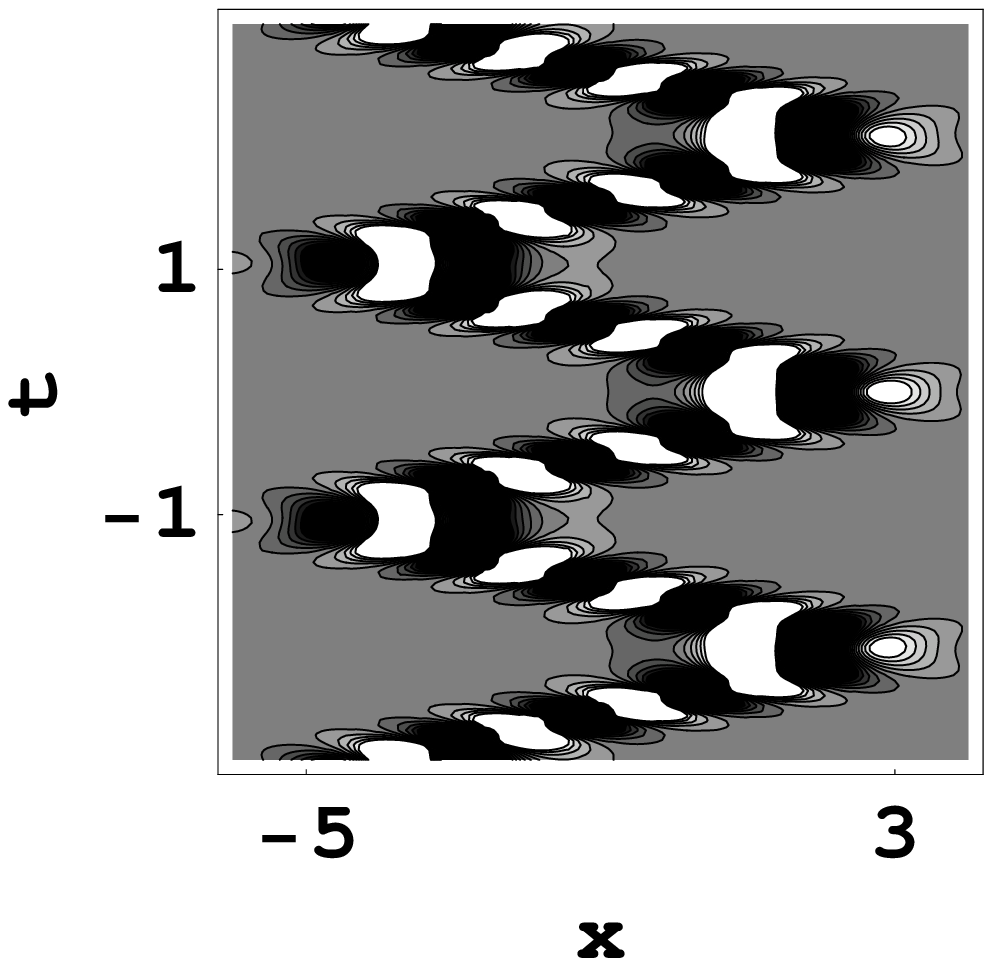}
\vspace{-0.5cm}{\center\hspace{3.3cm}\footnotesize ($a$)
\hspace{7.9cm}($b$)}\figcaption{(a) Breather propagation with
spectral parameters $\overline{\lambda_1}=1-i$,
$\overline{\lambda_2}=1+i$, initial phases $\xi_0=\xi_1=1$ and
coefficients $b(t)=Sin(3t)$, $c_0(t)=c_1(t)=0$ and $d(t)=Sin(10t)$;
(b) Contour plot of (a)}\label{breatherp}
\end{minipage}
\\[\intextsep]

Fig.~\ref{breatherp} shows a kind of periodic soliton solution for
the existence of two periodic coefficient $b(t)=Sin(3t)$ and
$d(t)=Sin(10t)$ supporting the periodically oscillating breather.
From the analytical discussion, the character line is determined
only by $b(t)$ when the other coefficients and parameters are given,
and the time dependent $d(t)$ gives rise to the periodical soliton
amplitude. In a conclusion, there exist three kinds of periodism
with respect to breather, character line and the soliton amplitude.

\begin{minipage}{\textwidth}
\renewcommand{\captionfont}{\scriptsize}
\renewcommand{\captionlabelfont}{\scriptsize}
\renewcommand{\captionlabeldelim}{.\,}
\renewcommand{\figurename}{Fig.\,}
\hspace{0.5cm}\includegraphics[scale=0.65]{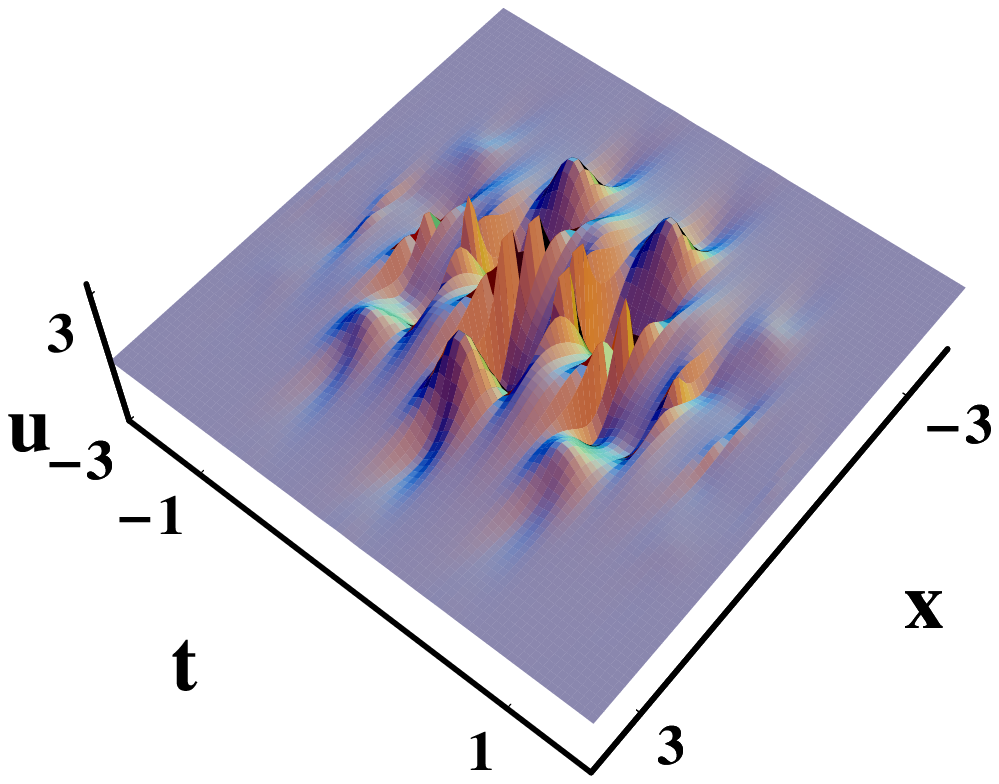}
\hspace{1.5cm}\includegraphics[scale=0.55]{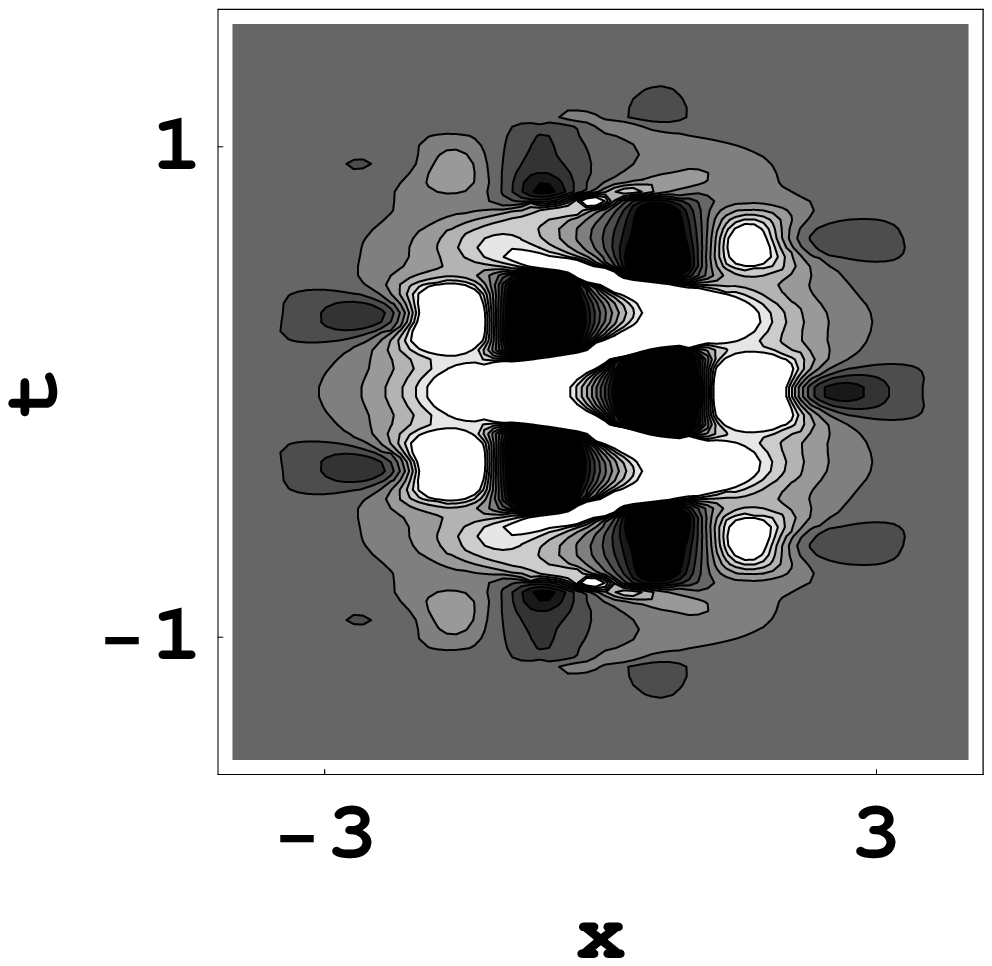}
\vspace{-0.5cm}{\center\hspace{3.3cm}\footnotesize ($a$)
\hspace{7.9cm}($b$)}\figcaption{(a) Interaction between a breather
and a soliton $\overline{\lambda_1}=1-i$,
$\overline{\lambda_2}=1+i$, $\overline{\lambda_3}=1$,
$\xi_1=\xi_2=\xi_3=0$, $b(t)=sin(10t)$, $c_0(t)=c_1(t)=0$ and
$d(t)=5\,t$, (b) Contour plot of (a)}\label{breather2}
\end{minipage}
\\[\intextsep]

The contour plot in Fig.~\ref{breather2} (b) exhibits the effects of
the dispersive term $b(t)$ and line-damping term $d(t)$ on the
interaction between a breather and a elevation. Being interacted by
breather propagation, the elevation undergoes a similar periodicity
corresponding to the breather and propagate without interrupting the
pecks of breather.

\begin{minipage}{\textwidth}
\renewcommand{\captionfont}{\scriptsize}
\renewcommand{\captionlabelfont}{\scriptsize}
\renewcommand{\captionlabeldelim}{.\,}
\renewcommand{\figurename}{Fig.\,}
\hspace{0.5cm}\includegraphics[scale=0.65]{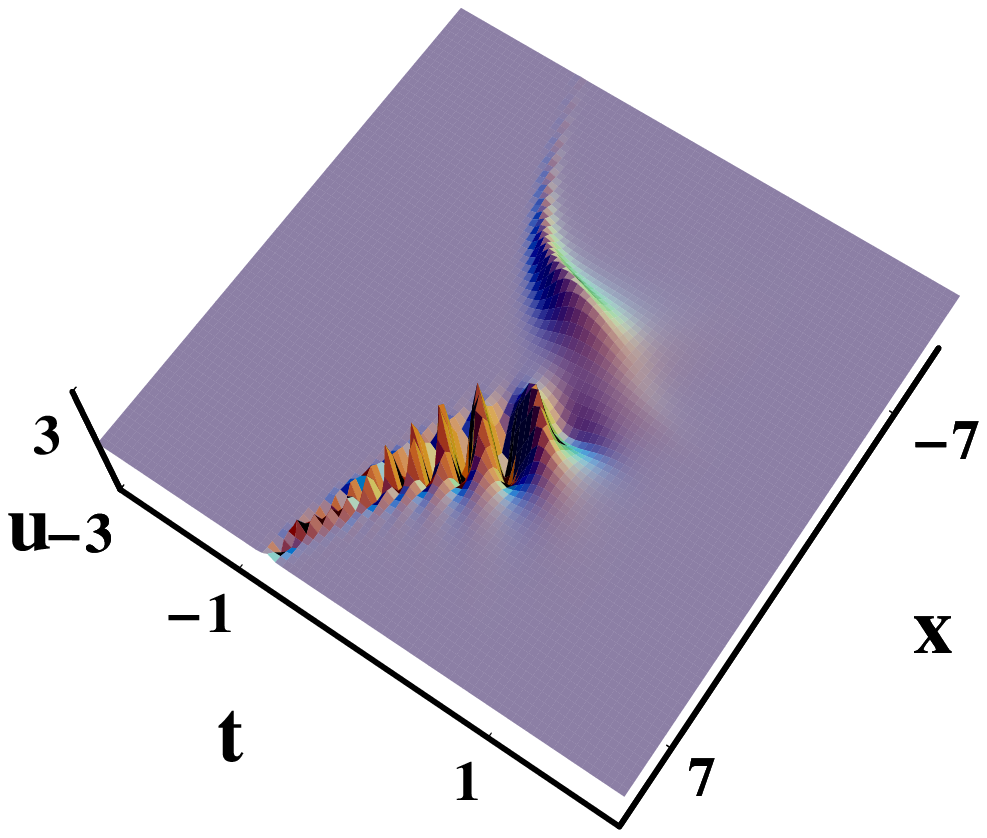}
\hspace{1.5cm}\includegraphics[scale=0.55]{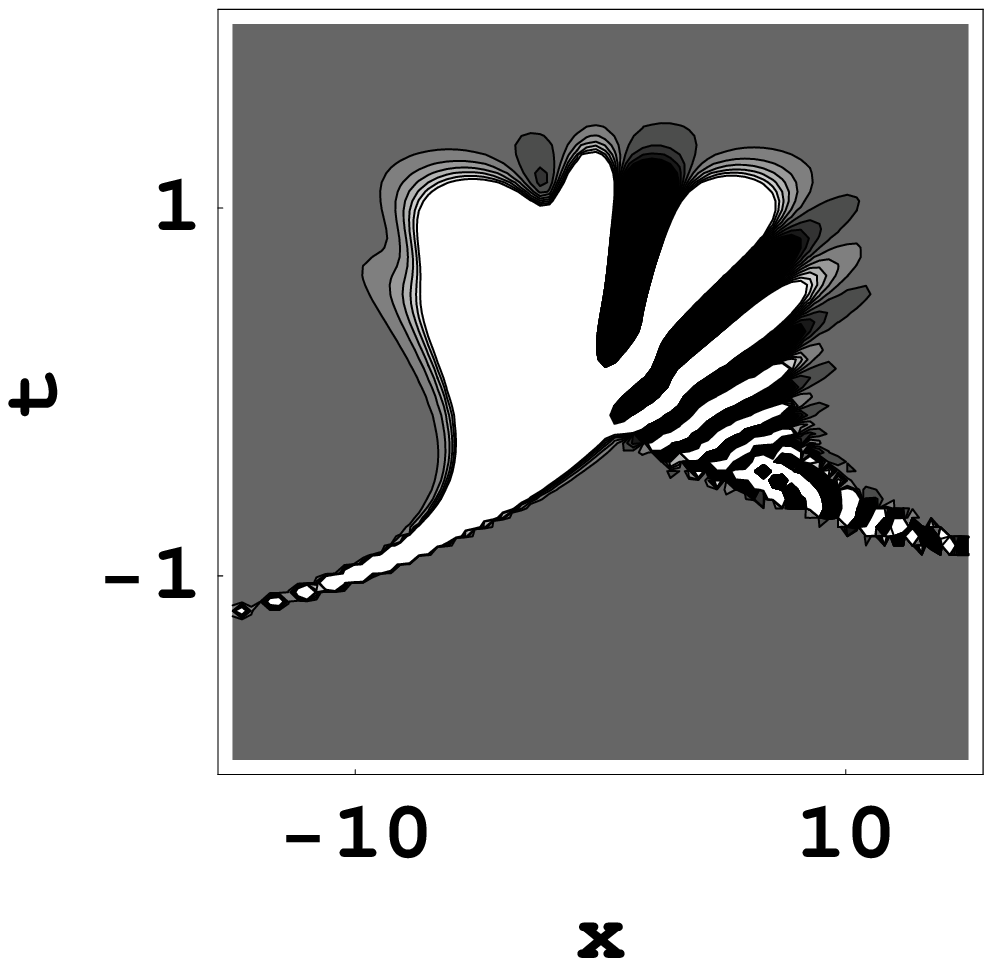}
\vspace{-0.5cm}{\center\hspace{3.3cm}\footnotesize ($a$)
\hspace{7.9cm}($b$)}\figcaption{(a) Interaction between a breather
and a soliton $\overline{\lambda_1}=1-i$,
$\overline{\lambda_2}=1+i$, $\overline{\lambda_3}=1$,
$\xi_1=\xi_2=\xi_3=0$, $b(t)=1$, $c_0(t)=0$, $c_1(t)=1$ and
$d(t)=10\,t$; (b) Contour plot of (a)}\label{breather3}
\end{minipage}
\\[\intextsep]

Fig.~\ref{breather3} depicts the collision between a soliton and
breather with approximately x=0 being the intersection point, which
witnesses the mutual interaction. The character line is overlapped
and swell phenomena occur corresponding to the soliton and breather
width. Compared with Fig.~\ref{breather2}, without periodical
time-varying function $b(t)$, Fig.~\ref{breather3} demonstrates
another kind of character line.

\begin{minipage}{\textwidth}
\renewcommand{\captionfont}{\scriptsize}
\renewcommand{\captionlabelfont}{\scriptsize}
\renewcommand{\captionlabeldelim}{.\,}
\renewcommand{\figurename}{Fig.\,}
\hspace{0.5cm}\includegraphics[scale=0.65]{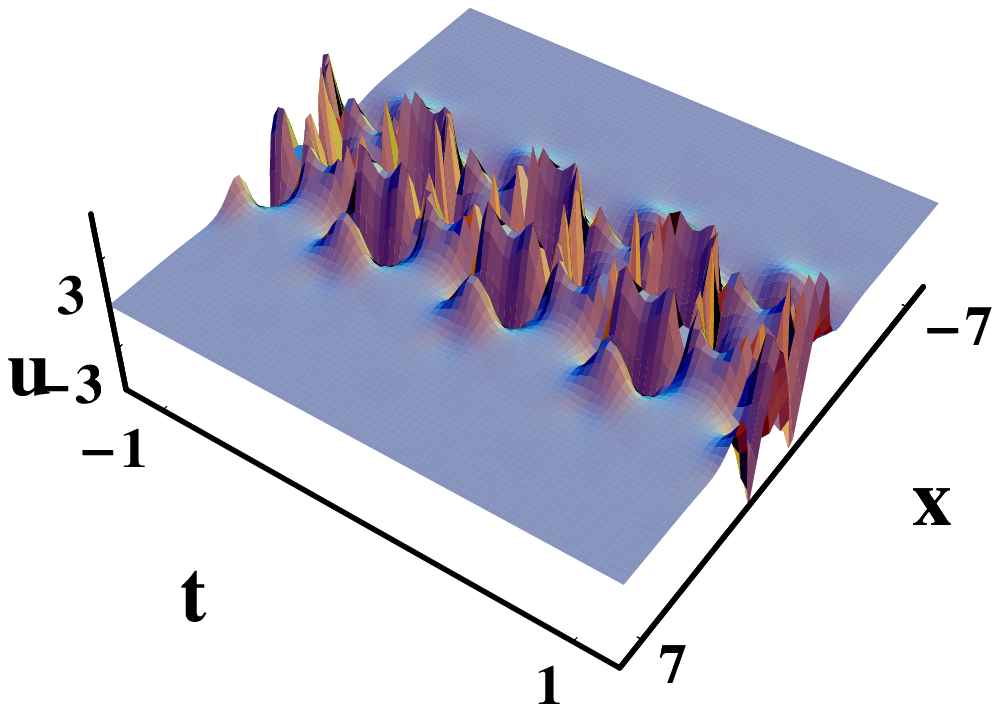}
\hspace{1.5cm}\includegraphics[scale=0.55]{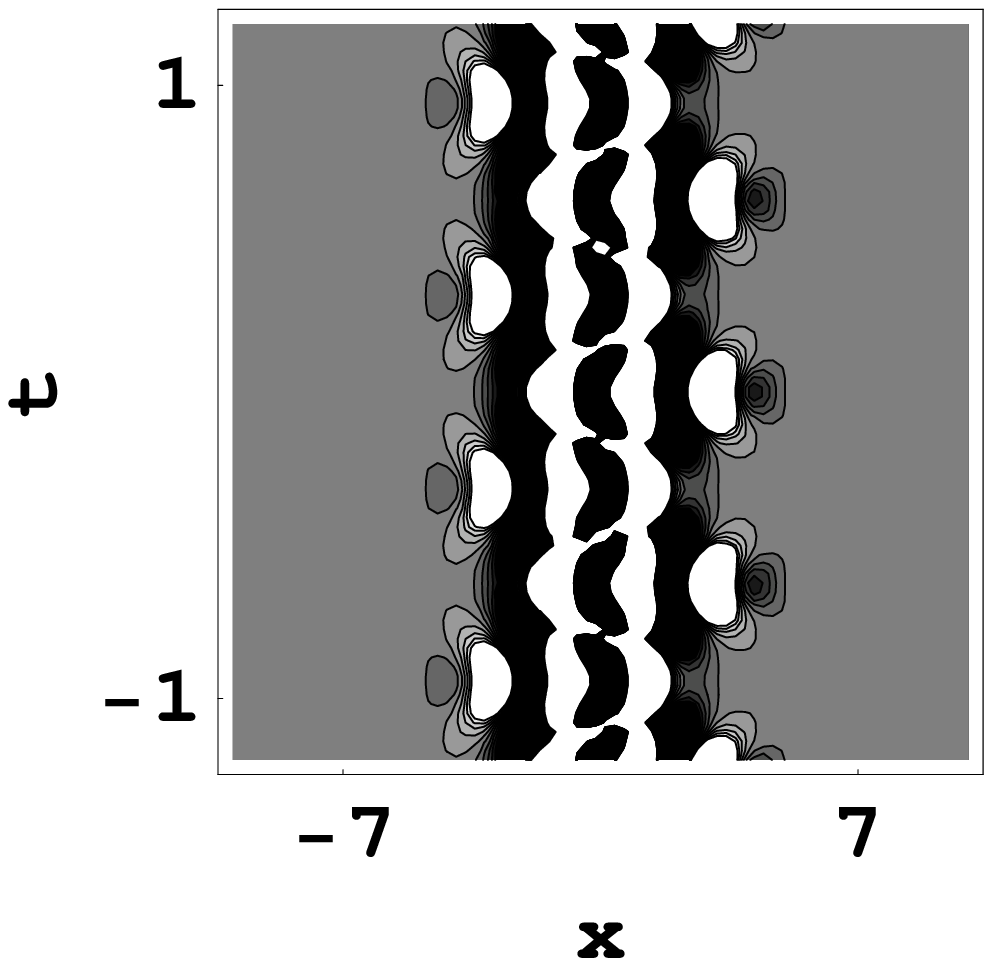}
\vspace{-0.5cm}{\center\hspace{3.3cm}\footnotesize ($a$)
\hspace{7.9cm}($b$)}\figcaption{(a) Interaction between two breather
with spectral parameters $\overline{\lambda_1}=1-i$,
$\overline{\lambda_2}=1+i$, $\overline{\lambda_3}=1.2-i$,
$\overline{\lambda_4}=1.2+i$ and $\xi_1=\xi_2=\xi_3=0$ and
coefficients $b(t)=sin(10t)$, $c_0(t)=c_1(t)=0$ and
$d(t)=5Sin(10t)$; (b) Contour plot of (a)}\label{breathersin}
\end{minipage}
\\[\intextsep]

With b(t) being periodical, we can observe two-breather periodical
oscillation in Fig.~\ref{breathersin}. The continuous interaction
between two breathers in one period is demonstrated in the contour
plot, which indicates the two-breather structure in this case can be
regarded as a coherent structure.

\begin{minipage}{\textwidth}
\renewcommand{\captionfont}{\scriptsize}
\renewcommand{\captionlabelfont}{\scriptsize}
\renewcommand{\captionlabeldelim}{.\,}
\renewcommand{\figurename}{Fig.\,}
\hspace{0.5cm}\includegraphics[scale=0.65]{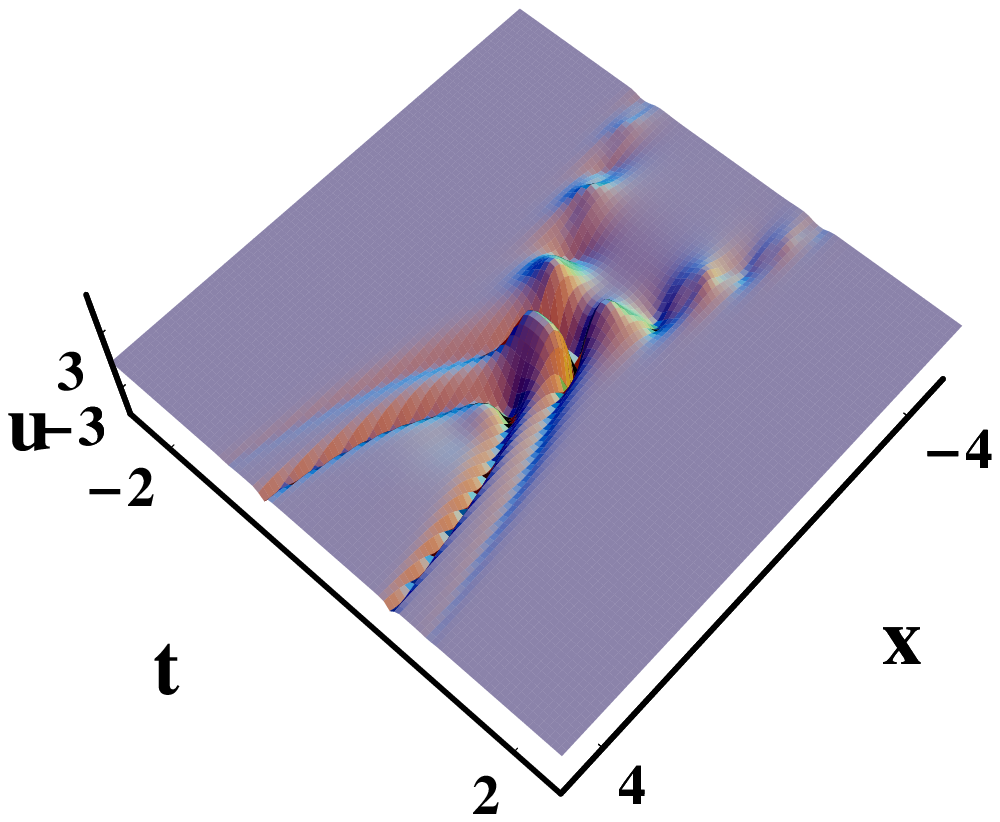}
\hspace{1.5cm}\includegraphics[scale=0.55]{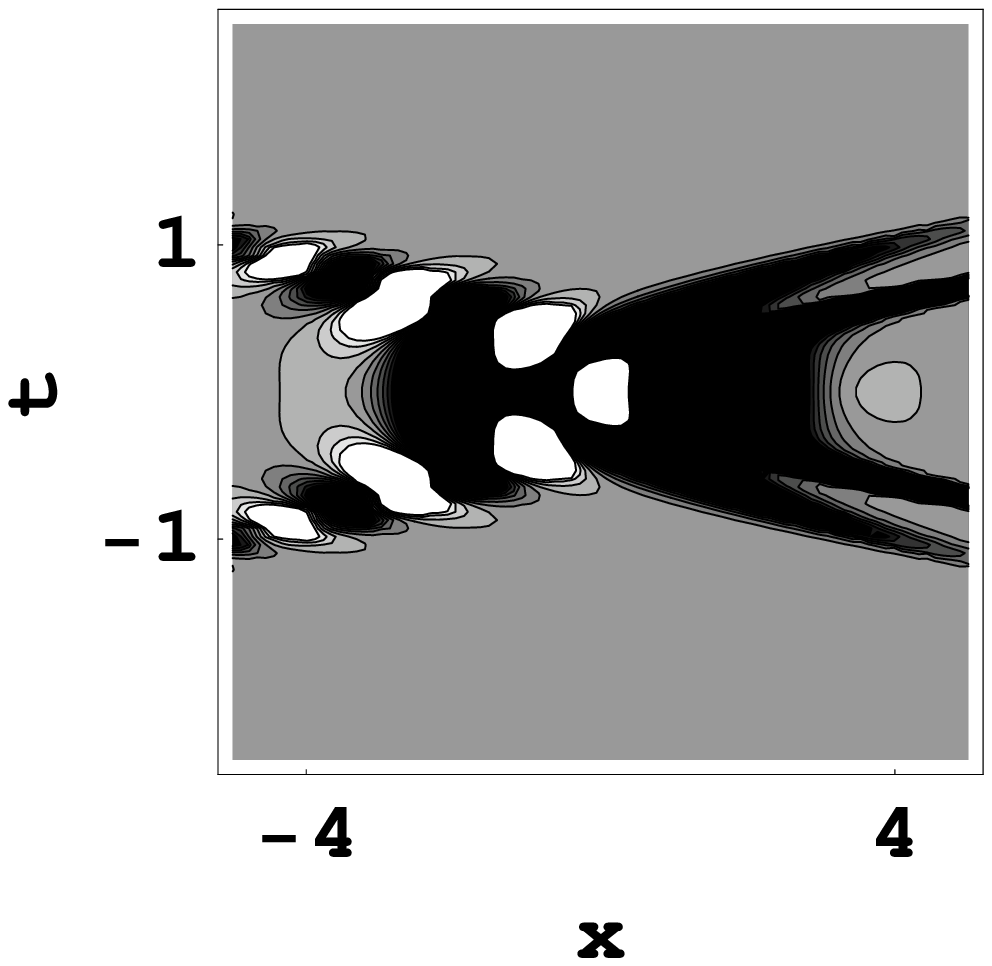}
\vspace{-0.5cm}{\center\hspace{3.3cm}\footnotesize ($a$)
\hspace{7.9cm}($b$)}\figcaption{(a) Interaction between a breather
and two solitons with spectral parameters
$\overline{\lambda_1}=1-i$, $\overline{\lambda_2}=1+i$,
$\overline{\lambda_3}=1.2$, $\overline{\lambda_4}=-2$ and
$\xi_1=\xi_2=\xi_3=\xi_4=0$ and coefficients $b(t)=t$,
$c_0(t)=c_1(t)=0$ and $d(t)=5t$; (b) Contour plot of
(a)}\label{breathert5t}
\end{minipage}
\\[\intextsep]

Fig.~\ref{breathert5t} illustrates the interaction between a
breather and two solitons. It is worth noting that with the
breather-soliton convergence at time approaching to zero, soliton
amplitude also reaches its peak value due to the line-damping term
$d(t)$. Compared with the structure including a soliton and a
breather in Figs.~\ref{breather2} and~\ref{breather3},
Fig.~\ref{breathert5t} presents three character lines, which
individually support two solitons and a breather in its propagation.

 \vspace{7mm} \noindent {\Large{\bf VII. Conclusions}}

\vspace{3mm} Based on the introduced constraint~(\ref{relation}),
under which Eq.~(\ref{equation}) is integrable and $c_1(t)$ becomes
independent from $d(t)$, the work of our paper can be concluded as
following:

(1) The Lax pair is generated by means of AKNS process with a
nonisospectral flow under constraint~(\ref{relation}). Meanwhile, DT
is constructed based on the Lax pair, by which multi-soliton and
breather solutions could be iterated from the seed ones.

(2) Analytical discussion corresponding to the characteristic line
is applied, in which the one soliton amplitude (polarity), width
(wave number) and velocity can be obtained. Meanwhile, with the aid
of one soliton solution, multi-soliton solutions are generated with
their graphical illustration. Initial phases are discussed by
employing a value shift, which gives rise to the soliton inverse
polarity. With appropriate selection of coefficients and spectral
parameters, such reverse polarity might prevent soliton cross. In
addition, variable coefficients influencing soliton amplitude,
velocity and width are investigated and the interaction among three
solitons is demonstrated.

(3) N-th iteration by DT can generate N-soliton solutions, including
the breathers in the periodically pulsating and isolated wave forms,
which are generated by employing a pair of complex conjugate
spectral parameters
($\overline{\lambda_1}=\alpha+i\beta,\overline{\lambda_2}=\alpha-i\beta$)
during iteration. For example, the four-soliton solution can
generate two breathers with periodical oscillation as shown in
Fig.~\ref{breathersin} or a breather and two solitons with three
character lines as shown in Fig.~\ref{breathert5t}. The inherent
periodism of breathers and the variable coefficients of
Eq.~(\ref{equation}) have coupling effects, which can yield abundant
structures of breather and soliton solutions, such as a cyclical
breather of three kinds of periodism, a breather-soliton interaction
with time-space locality and two-breathers in a coherent structure.
Such results can be extended to multi-soliton solutions by DT and
similar phenomena could be observed.

\vspace{7mm} \noindent {\Large{\bf Acknowledgements}}

\vspace{3mm}We express our sincere thanks to all the  members of our
discussion group for their valuable comments. This work has been
supported by the National Natural Science Foundation of China under
Grant No. 11302014, and by the Fundamental Research Funds for the
Central Universities under Grant Nos. 50100002013105026 and
50100002015105032 (Beijing University of Aeronautics and
Astronautics).

\end{document}